\documentclass[a4paper,twoside]{article}

\usepackage{epsfig}
\usepackage{subfigure}
\usepackage{calc}
\usepackage{amssymb}
\usepackage{amstext}
\usepackage{amsmath}
\usepackage{amsthm}
\usepackage{multicol}
\usepackage{pslatex}
\usepackage{apalike}

\usepackage{tikz-cd}
\usetikzlibrary{matrix,arrows}

\usepackage{graphicx}

\usepackage{SCITEPRESS}     

\begin{document}

\title{From Temporal Models to Property-Based Testing}

\author{\authorname{Nasser Alzahrani, Maria Spichkova and Jan Olaf Blech}
\affiliation{RMIT University, Melbourne, Australia}
\email{s3297335@student.rmit.edu.au, \{maria.spichkova,janolaf.blech\}@rmit.edu.au} }


\abstract{This paper presents a framework to apply property-based testing (PBT) on top of temporal formal models. The aim of this work is to help software engineers to understand temporal models that are presented formally and to make use of the advantages of formal methods: the core time-based constructs of a formal method are schematically translated to the BeSpaceD extension of the Scala programming language. This allows us to have an executable Scala code that corresponds to the formal model, as well as to perform PBT of the models functionality. To model temporal properties of the systems, in the current work we focus on two formal languages, TLA+ and Focus$^{ST}$.\footnote{Preprint. Accepted to the 12th International Conference on Evaluation of Novel Approaches to Software
		Engineering (ENASE 2017). Final version published by SCITEPRESS, http://www.scitepress.org}}

\onecolumn \maketitle \normalsize \vfill

\newcommand{\epar}[1]{``#1''}
\newcommand{\focust}{\textsc{Focus}$^{ST}$}

\section{\uppercase{Introduction}}
\label{sec:introduction}

Safety-critical systems, e.g., in the automotive domain \cite{efts_book}, become more and more software-intensive with every year.
While specifying such systems, a precise formal model, i.e., a mathematical model at some level of abstraction, might be essential to eliminate ambiguity and to detect possible errors early in the software development life-cycle (SDL). Also, in most cases the system properties have to be analysed in relation to the time, thus, verification/testing of the temporal aspects is crucial.

To achieve the integration of formal models into SDL, the development process should be human-oriented. Thus, aspects of human factors engineering should be taken into account, cf.  \cite{ICSE_2015_HF}. 
Moreover, using Formal Methods (FMs) can be beneficial  while developing not only safety-critical systems, but also web services, cf. \cite{Newcombe2015}.  
FMs were successfully applied to design and analyse systems since many years, cf. \cite{bowen1995seven,yu1999model}.
Despite all the advantages of FMs, software engineers are not keen to include them into the software development process. 
This problem was discussed 15-20  years ago, e.g., in \cite{hinchey_confessions_2003}. This problem is still unsolved now. 
Lack of readability and usability is one of the reasons for very limited use of FMs   in industrial projects   \cite{enase2016cfm}. 
However,  in some cases even simply implementable improvements can make an FM more readable and understandable, cf. 
  \cite{hffm_spichkova}.

In many cases, FMs require huge amount of training, as they use a very specific syntax that is unreadable for novices. 
In general, testing approaches are perceived by practitioners as more appropriate for a real-life development process.
However, they are usually comfortable with concepts from property-based testing (PBT), which require a little bit of mathematical thinking. 
PBT approach allows to use randomly generated test cases based on properties to test systems against their specifications.

To led programmers in formulating and testing properties of programs, Claessen and Hughes introduced a tool named \emph{QuickCheck} that is focusing on Haskell programming language. They demonstrated that \emph{QuickCheck}
allowed them to discover 
hundreds of bugs, e.g., DropBox file sharing service \cite{claessen_quickcheck:_2011,hughes2010software}. 
In its first edition, \emph{QuickCheck} was proposed as a testing framework for testing only functional programs. However, recent development in the area of PBT incorporates the state-fulness of systems. That provides functionality for the testing of state-ful systems as well as for testing programs written in 
imperative languages, e.g., C  \cite{gerdes_linking_2015,hughes2010software}.

We propose to apply PBT on top of temporal formal models. This might help software engineers to understand temporal formal models (which describe the state of a system in every discrete time point), as the FM constructs will be expressed in terms of system code.  This might contribute to the understandability of FMs indirectly, and allow software engineers to make use of the advantages of FMs. To achieve this goal, we suggest to translate the core time-based constructs of an FM  to the BeSpaceD extension of the Scala programming language, specified in \cite{blech_bespaced:_2014}. This allows us to have an executable Scala code that corresponds to the formal model, as well as to perform PBT of the models functionality. To model temporal properties of the systems, in the current work we focus on two formal languages, Temporal logic of actions (TLA+) and \focust. 
 TLA+ combines temporal logic with a logic of actions, and is used to describe behaviours of concurrent systems, cf. \cite{lamport_temporal_1994,lamport_hybrid_1993}.
  \focust is a formal language providing  concise but easily understandable specifications that is focused on  timing and spatial aspects of the system behaviour \cite{spichkova2014modeling,spichkova2007specification}.
 
To implement the proposed ideas, we selected Scala programming language, as on the PBT level this allows us to apply an extension to \emph{ScalaCheck library}. 
Early ideas of this approach was presented at 
 Software Technologies: Applications and Foundations Conference, cf. 
 \cite{Alzahrani2016}. 
In this paper we go further and discuss the developed framework and how it can be applies to TLA+ and \focust.
This approach is based on a completed Minor Master Thesis  of the first author.


\section{\uppercase{Proposed Framework}}
\label{sec:proposed-framework}

\begin{figure}[!h]
  \centering
   \includegraphics[scale=0.36]{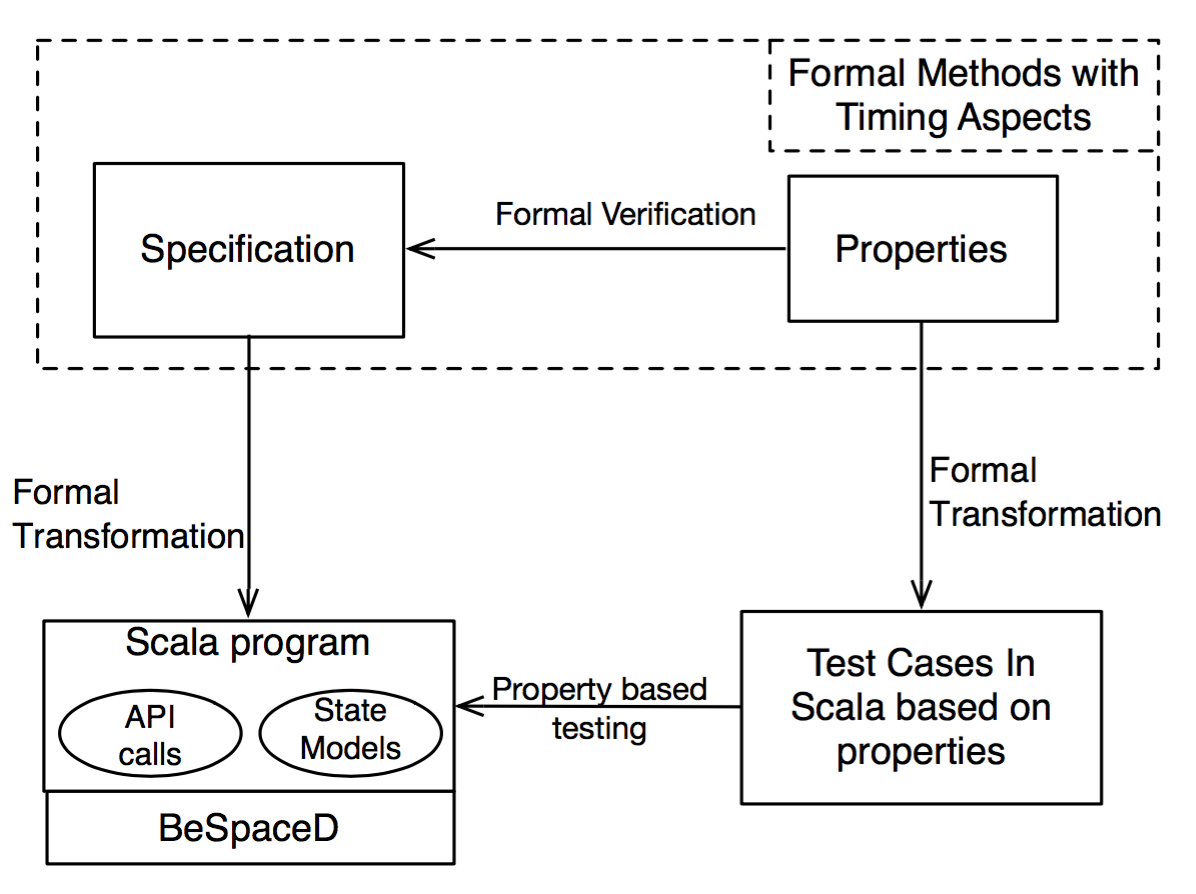}
  \caption{Proposed Framework}
  \label{fig:fm-general}
\end{figure}

Figure \ref{fig:fm-general} depicts the proposed framework that will allow for combining FMs with PBT. 
The general idea is to start with specifying the system using human-oriented modelling techniques based on FMs. 
After the specification phase, the software of the system under test is designed according to the specification.  The framework will then generate random test cases to exercise  and verify that the system
runs according to the specification. If a test fails, it will be the judgment of the engineer to decide whether
the errors were in the system software or in the specification formulas for which the system was not correctly specified. If the test passes without any errors, the system under test meets the specification. 

The FM specification gets translated to host programming language (Scala in this case). These specification gets formal verification depending on the flavour of FM being used. For example, in case of TLA+, the TLA+ model checker (TLC) is used to check the specification. 
On the other hand, in case of \focust, the theorem prover Isabelle/HOL via the framework \emph{Focus on Isabelle} 
is used to verify systems specification, cf. \cite{nipkow2002isabelle}  and \cite{spichkova2007specification}.

The workflow within the proposed framework includes the following steps:
\begin{itemize}
      \item To create an (informal) requirements specification of the system;
      \item To transform the informal specification to a formal specification (model) of the system, using  TLA+ or \focust;
      \item To verify formal model, using TLA+ model checker or Isabelle/HOL theorem prover, respectively;
      \item To translate the formal model to Scala using the provided translation schema;
      \item To add the specified in Scala model to the extended ScalaCheck library;
      \item  To check the extended ScalaCheck library against the behaviour generated by FM specification.
\end{itemize}

In this section, we show the applicability of the proposed framework to TLA+ and \focust. The goal is to demonstrate how the proposed framework can be applied to many types of FMs with similar syntax. Each subsection presents systematic informal program transformation schemas. Using these schemas makes transforming FM formulas to any hosting language, Scala in this case, an easy mechanical task. 
We start by analysing TLA+ syntax and semantics. After that, we show the design and model the API for the TLA+ flavour. After that,  we show the designed API and the testing it using small example (\emph{One Bit Block}). Similar process applied to \focust, showing the analysis of \focust syntax and semantics. Restricting \focust  to it's major parts that is related to temporal properties.

\subsection{Application to TLA+}

TLA provides a toolbox which includes an integrated development environment (IDE) for the TLA+.  
The IDE allows create and edit specifications, it also shows parsing errors and
can be used to turn TLA+ model checker. 
To decrease the cognitive load of the developer and tester, it also includes 
an error trace viewer and explorer: these components provide
a structured view of the states, 
illustrate how the states/values are changed at each step, and allow to run the TLA+ proof system.

A TLA formula such as $Init\land \square [Next]v$
specifies the initial states and the allowed transitions of a system.
It allows for transitions that do not change the value of \emph{v}. This kind of transitions is called \emph{stuttering transitions}.
Most TLA system specifications are of the form
$Init\land \square[Next]_{v} \land L$. The semantics of such formulas are shown in Table \ref{tab:tla-init}. 
 Table \ref{tab:tla-logic} shows logic operators in TLA+ and their mappings in Scala, many of the logical operators in Scala are provided by BeSpaceD.

\begin{table}[htp]
\caption{Semantics of TLA formula}
\begin{center}
{\footnotesize{
\begin{tabular}{l|p{60mm}}
  \hline
  Init & State formula describing the initial state(s)\\
  \hline
      Next & Action formula formalizing the transition relation --  usually a disjunction $A1 \lor..\lor An$ of possible actions (events) $Ai$ \\
  \hline
        L & Temporal formula asserting liveness conditions\\
  \hline
\end{tabular}
}}
\end{center}
\label{tab:tla-init}
\end{table}

\begin{table}[htp]
\caption{Operator mapping from TLA+ to Scala}
\begin{center}
{\footnotesize{
\begin{tabular}{c c}
\hline
TLA+ & Scala\\
\hline
\hline
 /\textbackslash & AND 
\\
\hline
 \textbackslash/ & OR 
\\
\hline
$\Rightarrow$ & IMPLIES
\\
\hline
TRUE & TRUE 
\\
\hline
 FALSE & FALSE
\\
\hline              
 BOOLEAN & Boolean 
\\
\hline              
$\left\{TRUE, FALSE\right\}$ & List(TRUE, FALSE) 
\\
\hline              
 $\leq$ & lessThanEq 
\\
\hline              
 $\geq$ & greaterThanEq 
\\
\hline              
 $>$ & greaterThan 
\\
\hline              
 $<$ & lessThan
\\
\hline              
 $\nleq$ & lessThanEqNot
\\
\hline              
 $\nless$ & lessThanNot
\\
\hline              
 $\ngeqslant$ & greaterThanEqNot
\\
\hline              
 $\ngtr$ & greaterThanNot
\\
\hline              
$\in$ & IN
\\
\hline
$x == e$ & defined(x, e)
\\
\hline
$x = e$ & assign(x, e)
\\
\hline
$\forall x \in S : p$ & for $\{ x \leftarrow S;$ if p$\}$ yield x
\\
\hline
$\exists x \in S : p$ & exists(x, S, p)
\\
\hline
CHOOSE $x \in S$ & choose(x, List(S))
\\
\hline
\end{tabular}
}}
\end{center}
\label{tab:tla-logic}
\end{table}

~\\
In TLA+, a representation of an abstraction of a system is modelled using the standard model. The Standard Model states that an abstract system is described as a collection of behaviours, each representing a possible execution of the system, where a behaviour is a sequence of states and a state is an assignment of values to variables.
In this model, an event (step) is the transition from one state to the next in a behaviour. 
For example, In one-bit clock, formulas are defined as follows:

{\footnotesize{
\begin{verbatim}
VARIABLE b 
Init ==  (b = 0) \/ (b = 1) 
Next == \/ /\ b = 0
           /\ b' = 1
        \/ /\ b = 1
           /\ b' = 0
\end{verbatim}
}}

\noindent
These two TLA+ statements define \emph{Init} and \emph{Next} to be  two formulas. Therefore, referencing  \emph{init} or \emph{Next} is completely equivalent to typing ((b = 0) \textbackslash/ (b = 1)). The equality symbol =  (typed ==) is read \emph{is defined to equal}. 
To transform these formulas into a host programming language, 
it is necessary to capture the essential aspects of the formula to be transformed, i.e., to create a translation schema. Each transformation step will consist of two elements: one to capture the TLA+ formula and one to capture the corresponding programming language function. The two schemata together can then be used to do the transformation. 
The TLA+ elements for the above formulas:

{\footnotesize{
\begin{verbatim}
f1 == p \/ q
f2 ==  \/ /\ p
          /\ q
       \/ /\ q
          /\ p
\end{verbatim}
}}

\noindent
That is, f1 represent Init, f2 represent Next, p represent (b=0), q represent (b=1) respectively.  
According to the translation schema, the translation of one bit clock from TLA+ to Scala is as follows:

{{\scriptsize{
\begin{verbatim}
val b: TLAVariable = TLAVariable(IN(List(0, 1)))
val init: TLAInit =  OR(defined(b,0), defined(b,1))
val next: TLANext = {
    while(true) {
        if defined(b, 0) 
           return assign(b, 1) 
        else 
          return assign(b, 0)
     }
}          
\end{verbatim}
}}}

\subsection{Application to \focust} 
\label{sec:app-focus}

The \focust language was inspired by Focus, a framework for formal specification and development of interactive systems. 
In both languages, specifications are based on the notion of streams, cf.  \cite{focus}. 
The syntax of \focust is particularly devoted to
specify spatial (S) and timing (T) aspects in a comprehensible fashion, which is the reason to extend the name of the language by ST: 
\focust stream is a mapping from natural numbers
to lists of messages within the corresponding time intervals.
Table \ref{tab:focus-op} shows a partial mappings between \focust basic operators and their Scala representations.

\begin{table}[htp]
\caption{Operator mapping from \focust to Scala}
\begin{center}
{\footnotesize{
\begin{tabular}{c c}
\hline
\focust & Scala\\
\hline
\hline
 /\textbackslash & AND 
\\
\hline
 \textbackslash/ & OR 
\\
\hline
$\to$ & IMPLIES
\\
\hline
TRUE & TRUE 
\\
\hline
 FALSE & FALSE
\\
\hline              
 BOOLEAN & Boolean 
\\
\hline              
 $\leq$ & lessThanEq 
\\
\hline              
 $\geq$ & greaterThanEq 
\\
\hline              
 $>$ & greaterThan 
\\
\hline              
 $<$ & lessThan
\\
\hline              
 $\nleq$ & lessThanEqNot
\\
\hline              
 $\nless$ & lessThanNot
\\
\hline              
 $\ngeqslant$ & greaterThanEqNot
\\
\hline              
 $\ngtr$ & greaterThanNot
\\
\hline              
$\in$ & IN
\\
\hline
$x == e$ & defined(x, e)
\\
\hline
$x = e$ & assign(x, e)
\\
\hline
$\forall x \in S : p$ & for $\{ x \leftarrow S;$ if p$\}$ yield x
\\
\hline
$\exists x \in S : p$ & exists(x, S, p)
\\
\hline
$\langle\rangle$  & List()
\\
\hline
$\langle a_1, \dots, a_m\rangle$ & $a_1$ to $a_m$
\\
\hline
\end{tabular}
}}
\end{center}
\label{tab:focus-op}
\end{table}

\begin{figure}[ht!]
  \centering
   \includegraphics[scale=0.4]{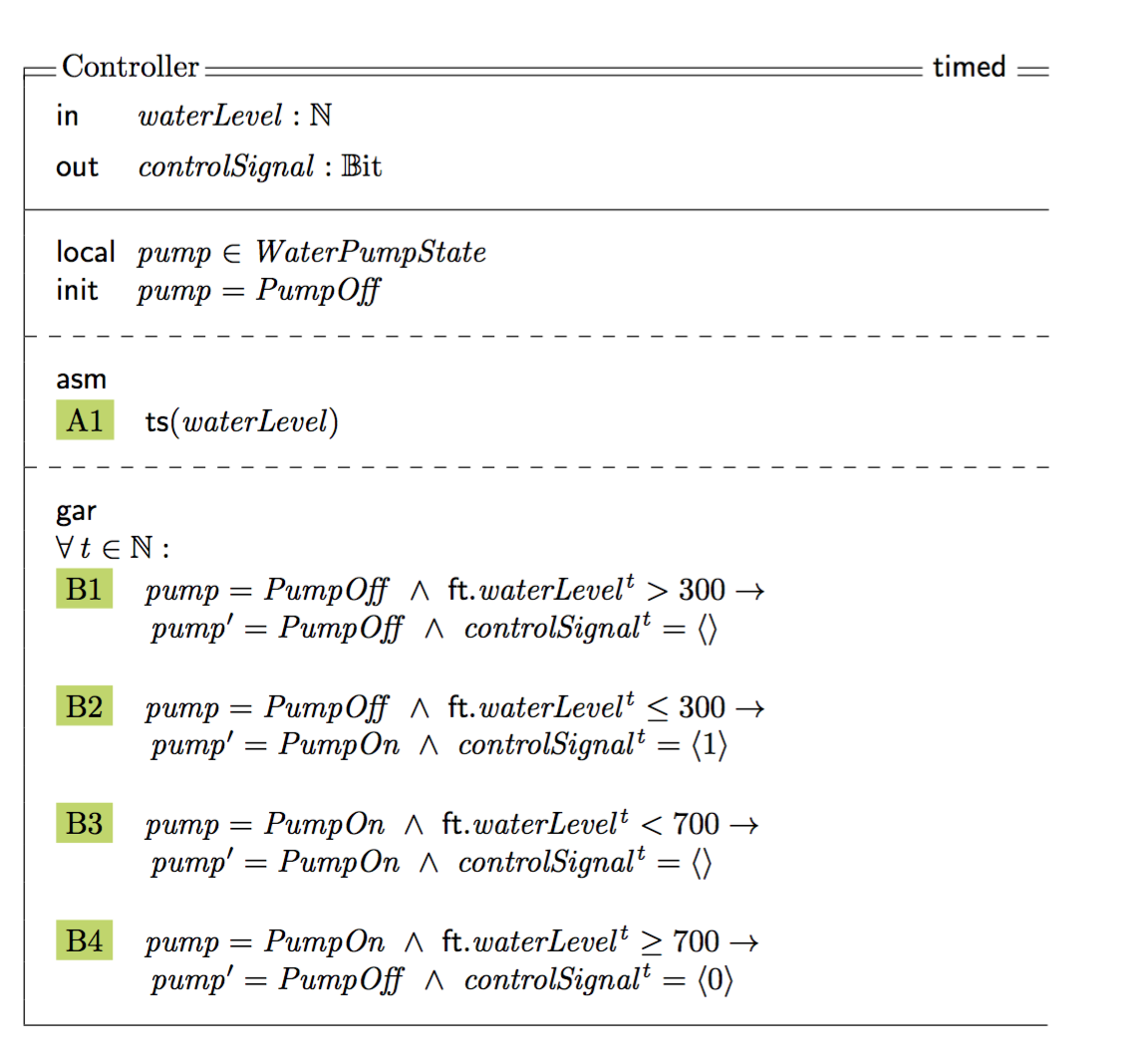}
  \caption{\focust Specification of Steam Boiler Controller \cite{spichkova2016spatio}}
  \label{fig:boiler}
\end{figure}

The \focust\ specification layout is based on human factor analysis within formal methods \cite{hffm_spichkova,Spichkova2013HFFM}.
Figure \ref{fig:boiler} provides an example on how a \focust specification looks like. 
The \emph{in} and \emph{out} sections of \focust specifications are used to specify input and output streams of the corresponding types.  \emph{local} and \emph{init} sections include local variables and initial values, respectively.  
\focust requires using assumption-guarantee templates, to avoid the omission of unnecessary assumptions about the system`s environment.
The keyword \emph{asm} lists the assumption that the specified component expect from its environment, e.g., the assumption $ts(s)$ would mean that the input stream $s$ should contain exactly one message per time interval.
The component behaviour that should be guaranteed in the case all assumptions are fulfilled, is then described in the specification section \emph{gar}.


\section{\uppercase{Discussion and Evaluation}}
\label{sec:evaluation}

Let us use the steam soiler \cite{focus} example to discuss the applicability of the developed framework. 
We selected this example as it  $(1)$ is simple-enough to introduce it shortly, $(2)$ is well-known example for analysing FMs, 
$(3)$ includes most of the functionalities of the proposed framework. 
For this example, we start by given the TLA+ and \focust specification which gets translated to Scala programming language before feeding the translated specification to the proposed framework. The translation correctness is verified manually by checking the behaviour that is generated by the tools developed and used to generate systems behaviours with the  behaviour generated by the actual TLA model checker (TLC).   
We follow the informal definition of the example provided in \cite{spichkova2016spatio}:
\emph{The steam boiler has a water tank, which contains a number of gallons of water, and
a pump, which adds $10$ gallons of water per time unit to its water tank, 
if the pump is on. At most $10$ gallons of water are consumed per time unit by
the steam production, if the pump is off.
The steam boiler has a sensor that measures the water level. 
Initially,  the water level is $500$ gallons, and the pump is off. 
In each time interval the system outputs it current water level in gallons
 and this level should always be between $200$ and $800$ gallons.}
 
The system consists of  three logical components: SteamBoiler, Converter, and Controller.
The specification \emph{Controller} as shown in Figure \ref{fig:boiler} describes the controller component of the system. The controller role is to switch the steam boiler pump on and off. In addition, it knows the current state of the pump. The behaviour of this component is asynchronous to keep the number of control signals as small as possible. 

\begin{figure}[!h]
  \centering
  \includegraphics[scale=0.42]{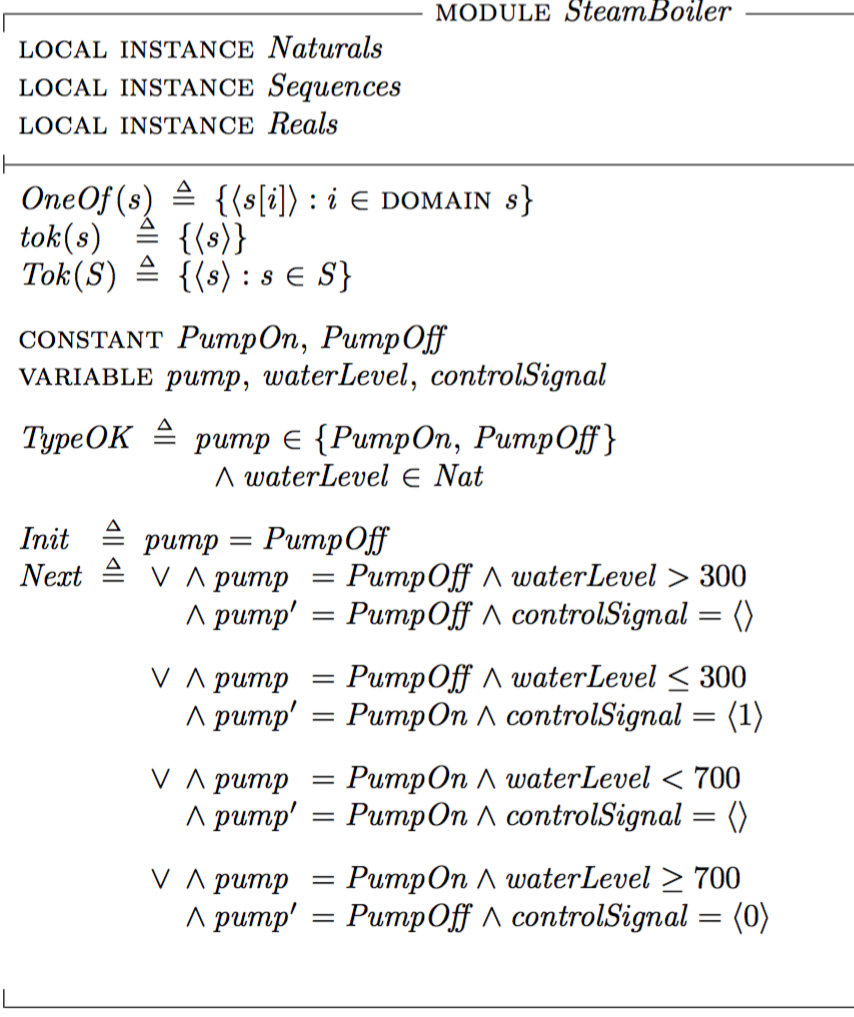}
  \caption{TLA+ specification of the Steam Boiler controller}
  \label{fig:boiler-tla}
\end{figure}

Figure \ref{fig:boiler-tla} shows TLA+ specification of the Steam Boiler controller. Unlike \focust,  TLA+ is  weakly typed. Therefore, it uses a convention to indicated types of variables using \emph{TypeOK} keyword as shown in the specification.

To check the framework, we provided two implementation for the steam boiler system, correct (wrt. the given \focust and TLA+ specification) and incorrect one (having mistakes wrt. the given specification). 
 For instance, in the case when the system is specified to have its current water level be between 200 and 800 gallons, the wrong implementation does not satisfy this property and instead have the the current level below 200 and above 800. The wrong example also include the failure of the pump to turn on or off. 
Table \ref{tab:invoc} shows number of invocations for every API call in each test run. 
Both translated TLA+ and \focust specifications have similar numbers since the schematic translation from TLA+ and \focust to Scala is similar in both cases.  
The extended ScalaCheck implementation that we developed does not shrink the test case to generate minimal failing test cases (which would make the code easier to debug). 
The future work will include the shrinking behaviour that is inspired by \emph{QuickCheck} library.

Table \ref{tab:perm-stat} contrast the performance of permutations and PBT test runs between the schematic translation of TLA+ and \focust. There are no observable differences between the performance of TLA+ and \focust in almost all of the phases of the workflow. This is expected since both TLA+ and \focust has similar syntax and the translation is similar in most cases. 
For the same reason, there is no considerable difference between lines of code after translation from TLA+ to Scala which was 70 lines of code and the translation from \focust to Scala was 75 lines. 
All tests were carried out on two machines: 

{\small
\begin{verbatim}
Intel Core i5 2.6 GHz,  RAM  8 GB 
Intel Core-i7 360QM 2.0 GHz, RAM  4GB   
\end{verbatim}
}

\begin{table}[ht!]
\caption{Number of API Invocations in test cases}
\begin{center}
\scalebox{0.8}{
\begin{tabular}{|c c c|}
\hline
 API Code &  TLA+  & \focust \\
\hline
\hline
\emph{startSystem()}  &  1 & 1 
\\
\hline
\emph{endSystem()}	  & 1 & 1 
\\
\hline
\emph{pumpDidOpen()} &  27 & 27
\\
\hline
\emph{openPump()}	  &  11 & 11
\\
\hline
\emph{pumpDidClose()} &  17 & 17
\\
\hline
\emph{closePump()}	  & 47 & 47 
\\
\hline
\emph{waterLevelDidChange(amount: Int)} &  21 & 21
\\
\hline
\emph{checkWaterLevel()}  &  20 & 20
\\
\hline
\emph{controlSignalDidChange(val: Int)}	  & 26 & 26
\\
\hline
\end{tabular}}
\end{center}
\label{tab:invoc}
\end{table}

\begin{table}[ht!]
\caption{Translated TLA+ and \focust statistics (time in seconds)}
\begin{center}
\scalebox{0.80}{
\begin{tabular}{|c c c|}
\hline
 & TLA+ & \focust\\
\hline
\hline
API permutations & 10-11  & 10-11
\\
\hline
Behaviour Generating & 7-8  &7-8 
\\
\hline
Single Test run   & 0.5 & 0.5 
\\
\hline
Total Test run time 100 test cases & 23-25 & 23-25 
\\
\hline

\end{tabular}}
\end{center}
\label{tab:perm-stat}
\end{table}

\noindent
To evaluate the performance of the scripts using to support the framework, we used a number of further problems commonly used in 
the TLA+ community:  
\begin{itemize}
\item
\emph{One Bit Clock} simply alternates between $0$ and $1$. Such a clock is used to control any modern computer. Its times being displayed as the voltage on a wire. Therefore, there are only two states; the \emph{0 state} and the \emph{1 state}. 
\item
The \emph{DieHard problem}  from the movie\emph{Die Hard 3}, the heroes had to solve the problem of obtaining exactly 4 gallons of water using a 5 gallon jug, a 3 gallon jug, and a water faucet. 
\item
Euclid`s algorithm  for computing the greatest common divisor of two positive integers. 
\item
Therac-25, a radiation therapy machine used in curing cancer,  
led to deaths and serious injuries of patients which received thousand times the normal dose of radiation \cite{Miller1987,leveson1993investigation}. 
The causes of these accidents were software failures as well as problems with the system interface. 
The machine included VT-100 terminal which controlled the PDP-11 computer, where the sequence of user actions leading to the accidents was as follows: 
 user selects 25 MeV photon mode, 
enters \emph{cursor up},
select 25 MeV Electron mode, previous commands have to take place in eight seconds.
\end{itemize}

\noindent
Table \ref{tab:tla-stat} shows the statistics on the applied behaviour generator.  
\emph{Diameter} column is the number of states in the longest path of the graph in which no state appears twice.
\emph{States Found} column is the total number of  states it examined in the first step of the algorithm or as successor states in the second step.
\emph{Distinct States} column is the number of states that form the set of nodes of the graph. For instance, in case of \emph{One Bit Clock}, model checker found two distinct states.

\begin{table}[htp]
\caption{Behaviour Generator Statistics}
\begin{center}
\scalebox{0.80}{
\begin{tabular}{|c c c c|}
\hline
Example & Diameter & State Found & Distinct States\\
\hline
\hline
DieHard & 9  & 97 & 16
\\
\hline
One Bit Clock  & 1  & 4 & 2
\\
\hline
Euclid Algorithm  & 3 & 22 & 8
\\
\hline
Therac25 & 9  & 97 & 16
\\
\hline

\end{tabular}}
\end{center}
\label{tab:tla-stat}
\end{table}

\section{\uppercase{Conclusion}}
\label{sec:conclusions}
 
We have presented our framework for application of the property-based testing (PBT) concepts on top of temporal formal models. 
This allows us to have an executable Scala code that corresponds to the formal model, as well as to perform PBT of the models functionality. 
The framework is aiming on  reduction of the impedance mismatch between formal methods and practitioners through the combining of formal methods with property-based testing. 
We introduced the core ideas on how the framework can be applied to particular formal languages, such as  TLA+ and Focus$^{ST}$.

\bibliographystyle{apalike}
{\small

\begin{thebibliography}{}

\bibitem[Alzahrani et~al., 2016]{Alzahrani2016}
Alzahrani, N., Spichkova, M., and Blech, J.~O. (2016).
\newblock {\em Spatio-Temporal Models for Formal Analysis and Property-Based
  Testing}, pages 196--206.
\newblock Springer.

\bibitem[Blech and Schmidt, 2014]{blech_bespaced:_2014}
Blech, J.~O. and Schmidt, H. (2014).
\newblock {BeSpaceD}: Towards a tool framework and methodology for the
  specification and verification of spatial behavior of distributed software
  component systems.
\newblock {\em CoRR}.

\bibitem[Bowen and Hinchey, 1995]{bowen1995seven}
Bowen, J.~P. and Hinchey, M.~G. (1995).
\newblock Seven more myths of formal methods.
\newblock {\em IEEE software}, 12(4):34.

\bibitem[Broy and St{\o}len, 2001]{focus}
Broy, M. and St{\o}len, K. (2001).
\newblock {\em Specification and Development of Interactive Systems: Focus on
  Streams, Interfaces, and Refinement}.
\newblock Springer.

\bibitem[Claessen and Hughes, 2011]{claessen_quickcheck:_2011}
Claessen, K. and Hughes, J. (2011).
\newblock {QuickCheck}: A lightweight tool for random testing of haskell
  programs.
\newblock {\em {SIGPLAN} Not.}, 46(4):53--64.

\bibitem[Gerdes et~al., 2015]{gerdes_linking_2015}
Gerdes, A., Hughes, J., Smallbone, N., and Wang, M. (2015).
\newblock Linking unit tests and properties.
\newblock In {\em {SIGPLAN} Workshop}, pages 19--26. {ACM}.

\bibitem[Hinchey, 2003]{hinchey_confessions_2003}
Hinchey, M.~G. (2003).
\newblock Confessions of a formal methodist.
\newblock In {\em Safety Critical Systems and Software}, pages 17--20. ACS.

\bibitem[Hughes, 2010]{hughes2010software}
Hughes, J. (2010).
\newblock Software testing with quickcheck.
\newblock In {\em Central European Functional Programming School}, pages
  183--223. Springer.

\bibitem[K{\"u}hnel and Spichkova, 2007]{efts_book}
K{\"u}hnel, C. and Spichkova, M. (2007).
\newblock Fault-tolerant communication for distributed embedded systems.
\newblock In {\em Software Engineering of Fault Tolerance Systems}, volume~19,
  page 175. World Scientific Publishing.

\bibitem[Lamport, 1993]{lamport_hybrid_1993}
Lamport, L. (1993).
\newblock Hybrid systems in {TLA}+.
\newblock In Grossman, R.~L., Nerode, A., Ravn, A.~P., and Rischel, H.,
  editors, {\em Hybrid Systems}, number 736 in LNCS, pages 77--102. Springer.

\bibitem[Lamport, 1994]{lamport_temporal_1994}
Lamport, L. (1994).
\newblock The temporal logic of actions.
\newblock 16(3):872--923.

\bibitem[Leveson and Turner, 1993]{leveson1993investigation}
Leveson, N.~G. and Turner, C.~S. (1993).
\newblock An investigation of the therac-25 accidents.
\newblock {\em Computer}, 26(7):18--41.

\bibitem[Miller, 1987]{Miller1987}
Miller, E. (1987).
\newblock {The Therac-25 Experience}.
\newblock In {\em Conf. State Radiation Control Program Directors}.

\bibitem[Newcombe et~al., 2015]{Newcombe2015}
Newcombe, C., Rath, T., Zhang, F., Munteanu, B., Brooker, M., and Deardeuff, M.
  (2015).
\newblock {How Amazon Web Services Uses Formal Methods}.
\newblock {\em CACM}, 58(4):66--73.

\bibitem[Nipkow et~al., 2002]{nipkow2002isabelle}
Nipkow, T., Paulson, L.~C., and Wenzel, M. (2002).
\newblock {\em {Isabelle/HOL: a proof assistant for higher-order logic}},
  volume 2283.
\newblock Springer Science \& Business Media.

\bibitem[Spichkova, 2007]{spichkova2007specification}
Spichkova, M. (2007).
\newblock {\em {Specification and seamless verification of embedded real-time
  systems: FOCUS on Isabelle}}.
\newblock PhD thesis, Technical University Munich.

\bibitem[Spichkova, 2012]{hffm_spichkova}
Spichkova, M. (2012).
\newblock {Human Factors of Formal Methods}.
\newblock In {\em IADIS Interfaces and Human Computer Interaction 2012}.

\bibitem[Spichkova, 2013]{Spichkova2013HFFM}
Spichkova, M. (2013).
\newblock {\em Design of formal languages and interfaces: ``Formal'' does not
  mean ``unreadable''}.
\newblock IGI Global.

\bibitem[Spichkova, 2016]{spichkova2016spatio}
Spichkova, M. (2016).
\newblock {Spatio-temporal features of Focus$^{ST}$}.
\newblock {\em CoRR}.

\bibitem[Spichkova et~al., 2014]{spichkova2014modeling}
Spichkova, M., Blech, J.~O., Herrmann, P., and Schmidt, H.~W. (2014).
\newblock {Modeling Spatial Aspects of Safety-Critical Systems with
  Focus$^{ST}$}.
\newblock In {\em MoDeVVa}, pages 49--58.

\bibitem[Spichkova et~al., 2015]{ICSE_2015_HF}
Spichkova, M., Liu, H., Laali, M., and Schmidt, H.~W. (2015).
\newblock Human factors in software reliability engineering.
\newblock {\em Workshop on Applications of Human Error Research to Improve
  Software Engineering}.

\bibitem[Yu et~al., 1999]{yu1999model}
Yu, Y., Manolios, P., and Lamport, L. (1999).
\newblock Model checking tla+ specifications.
\newblock In {\em Correct Hardware Design and Verification Methods}, pages
  54--66. Springer.

\bibitem[Zamansky et~al., 2016]{enase2016cfm}
Zamansky, A., Rodriguez-Navas, G., Adams, M., and Spichkova, M. (2016).
\newblock Formal methods in collaborative projects.
\newblock In {\em {11th International Conference on Evaluation of Novel
  Approaches to Software Engineering}}. IEEE.

\end{thebibliography}

}

\vfill
\end{document}